\documentclass[conference]{IEEEtran}
\IEEEoverridecommandlockouts

\usepackage{cite}
\usepackage{amsmath,amssymb,amsfonts}
\usepackage{algorithmic}
\usepackage{graphicx}
\usepackage{textcomp}
\usepackage{xcolor}
\usepackage{amsmath,amsfonts}
\usepackage{array}

\usepackage[bookmarks=false]{hyperref} 
\usepackage{float}
\usepackage{textcomp}
\usepackage{stfloats}
\usepackage{url}
\usepackage{verbatim}
\usepackage{graphicx}
\usepackage{placeins} 
\usepackage{pifont}
\usepackage{booktabs}
\usepackage{multirow}
\usepackage[linesnumbered,ruled,vlined]{algorithm2e}
\RequirePackage{letltxmacro}
\LetLtxMacro{\LaTeXtextbf}{\textbf}
\LetLtxMacro{\textbf}{\LaTeXtextbf}
\usepackage{cite}
\usepackage{amsmath,amssymb,amsfonts}
\usepackage{graphicx}
\usepackage{textcomp}
\usepackage{color}
\usepackage{caption}
\usepackage{subcaption}
\usepackage{enumitem}  
\def\BibTeX{{\rm B\kern-.05em{\sc i\kern-.025em b}\kern-.08em
    T\kern-.1667em\lower.7ex\hbox{E}\kern-.125emX}}

\usepackage{tcolorbox}
\tcbuselibrary{breakable}

\newcommand{\eg}{\textit{\eg,}}
\usepackage{rotating}
\usepackage{multirow}
\usepackage[affil-it]{authblk} 

\ifCLASSINFOpdf

\else

\fi

\begin{document}

\title{B5GRoam: A Zero Trust Framework for Secure and Efficient On-Chain B5G Roaming}



\author[1,2]{Mohamed Abdessamed Rezazi}
\author[1]{Mouhamed Amine Bouchiha}
\author[1]{Ahmed Mounsf Rafik Bendada}
\author[1]{\\Yacine Ghamri-Doudane}

\affil[1]{L3i, La Rochelle University, La Rochelle, France}
\affil[2]{Ecole Nationale Supérieure d'Informatique, Algiers, Algeria}

\maketitle

\begin{abstract}
Roaming settlement in 5G and beyond networks demands secure, efficient, and trustworthy mechanisms for billing reconciliation between mobile operators. While blockchain promises decentralization and auditability, existing solutions suffer from critical limitations—namely, data privacy risks, assumptions of mutual trust, and scalability bottlenecks. To address these challenges, we present B5GRoam, a novel on-chain and zero-trust framework for secure, privacy-preserving, and scalable roaming settlements. B5GRoam introduces a cryptographically verifiable call detail record (CDR) submission protocol, enabling smart contracts to authenticate usage claims without exposing sensitive data. To preserve privacy, we integrate non-interactive zero-knowledge proofs (zkSNARKs) that allow on-chain verification of roaming activity without revealing user or network details. To meet the high-throughput demands of 5G environments, B5GRoam leverages Layer 2 zk-Rollups, significantly reducing gas costs while maintaining the security guarantees of Layer 1. Experimental results demonstrate a throughput of over 7,200 tx/s with strong privacy and substantial cost savings. By eliminating intermediaries and enhancing verifiability, B5GRoam\footnote{\href{https://github.com/Brivan-26/B5GRoam/tree/main }{https://github.com/Brivan-26/B5GRoam/tree/main }}  offers a practical and secure foundation for decentralized roaming in future mobile networks.
\end{abstract}



\section{Introduction}


\IEEEPARstart{A}{s} we transition toward Beyond 5G (B5G) networks, the expectations placed on mobile infrastructure are evolving rapidly, not only in terms of performance but also in terms of flexibility, autonomy, and trust \cite{B5Gslicingsrv2024}. These networks must support increasingly heterogeneous actors and services, ranging from traditional Mobile Network Operators (MNOs) to emerging vertical-specific micro-operators. In such a diverse and dynamic environment, legacy roaming mechanisms struggle to provide the agility and transparency required for seamless inter-operator collaboration. This calls for a fundamental rethinking of how roaming is managed, favoring decentralized architectures that can meet B5G’s requirements for security and interoperability.

\quad The increasing complexity of mobile network ecosystems marked by the proliferation of local 5G operators (L5GOs), private enterprise networks, and dynamic service demands has highlighted significant limitations in traditional centralized roaming architectures. Existing systems heavily rely on intermediaries such as data and financial clearing houses to coordinate roaming agreements, process usage records, and manage settlements ~\cite{roaminfocom24}. These intermediaries introduce latency and cost overhead and also represent single points of failure and trust. Moreover, the lack of transparency and real-time responsiveness impedes fraud detection and resolution. In this evolving landscape, decentralizing roaming services presents a compelling alternative. This aims to enable automated, transparent, and tamper-resistant coordination between MNOs and other stakeholders without the need for centralized oversight \cite{nguyen2022blockroam}.

\quad Blockchain technology has emerged as a powerful tool for enabling decentralization, offering a tamper-proof and transparent method for recording transactions \cite{lbdt2024,diallo2024trade}. In the context of 5G roaming, blockchain has the potential to remove intermediaries, reduce fraud, and improve settlement efficiency. However, existing blockchain-based solutions in this domain remain limited. Many rely on public blockchains, which pose significant privacy risks by exposing sensitive data such as call detail records (CDRs) on-chain. Additionally, most frameworks assume trust in the participating entities, lacking sufficient mechanisms for verifying data authenticity or resolving disputes. Finally, scalability remains a critical challenge, as many blockchain systems struggle to handle the high transaction throughput needed for real-time 5G roaming settlements.

To overcome these limitations, we propose a novel blockchain-based roaming settlement framework that ensures verifiability, privacy, and scalability by design. Our approach introduces a verifiable CDRs submission protocol in which each record is cryptographically signed and auditable, enabling an on-chain smart contract to verify the authenticity of CDRs before initiating settlement. To preserve user privacy, we leverage zero-knowledge proofs (ZKPs) to validate usage claims without disclosing sensitive data on-chain. This prevents both data leakage and unauthorized surveillance. To support large-scale deployment, our system is built on a scalable Layer-2 architecture utilizing zkRollups, which reduces gas fees while preserving the integrity and finality of on-chain operations. Unlike prior work, our solution incorporates a trustless and privacy-preserving settlement mechanism and ensures compatibility with existing mobile network practices through modular integration points. This positions our framework as a practical and secure alternative for next-generation 5G roaming systems.


\section{Related work} \label{sec:relatedworks}

\quad 
In this section, we present and analyze the relevant contributions to on-chain 5G roaming management, highlighting the aspects of each solution and its limitations.

Recent research has explored the use of blockchain to decentralize 5G roaming management, offering improvements in efficiency, transparency, and fraud reduction. Refaey et al.~\cite{refaey2019policy} introduced a smart contract-driven policy and charging control framework enabling flexible, interoperable roaming among MNOs. While their EU-based case study demonstrated improved revenue and satisfaction, the solution lacked performance metrics and clarity on billing settlement. Similarly, Mafakheri et al.~\cite{mafakheri2021roaming} leveraged Hyperledger to enable automated billing, peer-to-peer payments, and identity management with enhanced privacy via private channels. However, they provided limited details on system implementation, payment logic, and safeguards against off-chain manipulation.

Other works address fraud and performance challenges with innovative designs but still leave critical gaps. Weerasinghe et al.~\cite{weerasinghe2021l5go} proposed a platform for L5G operators with a reputation system and dynamic roaming via smart contracts, but their system faced latency issues and lacked privacy protections. Nguyen et al. BlockRoam~\cite{nguyen2022blockroam} improved latency by using a fast PoS-based consensus and dynamic leader selection, but lacks finality and still exposes sensitive data publicly. Across these efforts, common limitations include insufficient privacy safeguards, a lack of trustless settlement mechanisms, and scalability concerns—gaps that motivate more robust, privacy-preserving, and verifiable blockchain-based roaming solutions.

While blockchain offers key properties for decentralizing and securing 5G roaming, existing solutions fall short in addressing foundational issues, namely CDR privacy, trustless verification, and scalability. This work is the first to propose a fully verifiable, privacy-preserving blockchain-based roaming framework that integrates zkSNARKs for secure CDR handling, a trustless proof submission protocol, and a Layer-2 zkRollup solution for scalability. Our design aims to eliminate unverifiable trust assumptions or expose sensitive user data while achieving practical on-chain and off-chain performance.

\section{B5GRoam Framework} \label{sec:sysArch}

\quad In this section, we present the B5GRoam framework, which builds privacy-preserving data handling mechanisms using zkSNARKs, a verifiable CDR submission protocol, and a Layer-2 rollup-based scalability solution for secure and efficient on-chain roaming management. We first present the system and threat models, and then provide a detailed description of the proposed framework and its workflow.

\begin{figure}[t]
\centering
\includegraphics[scale=0.16]{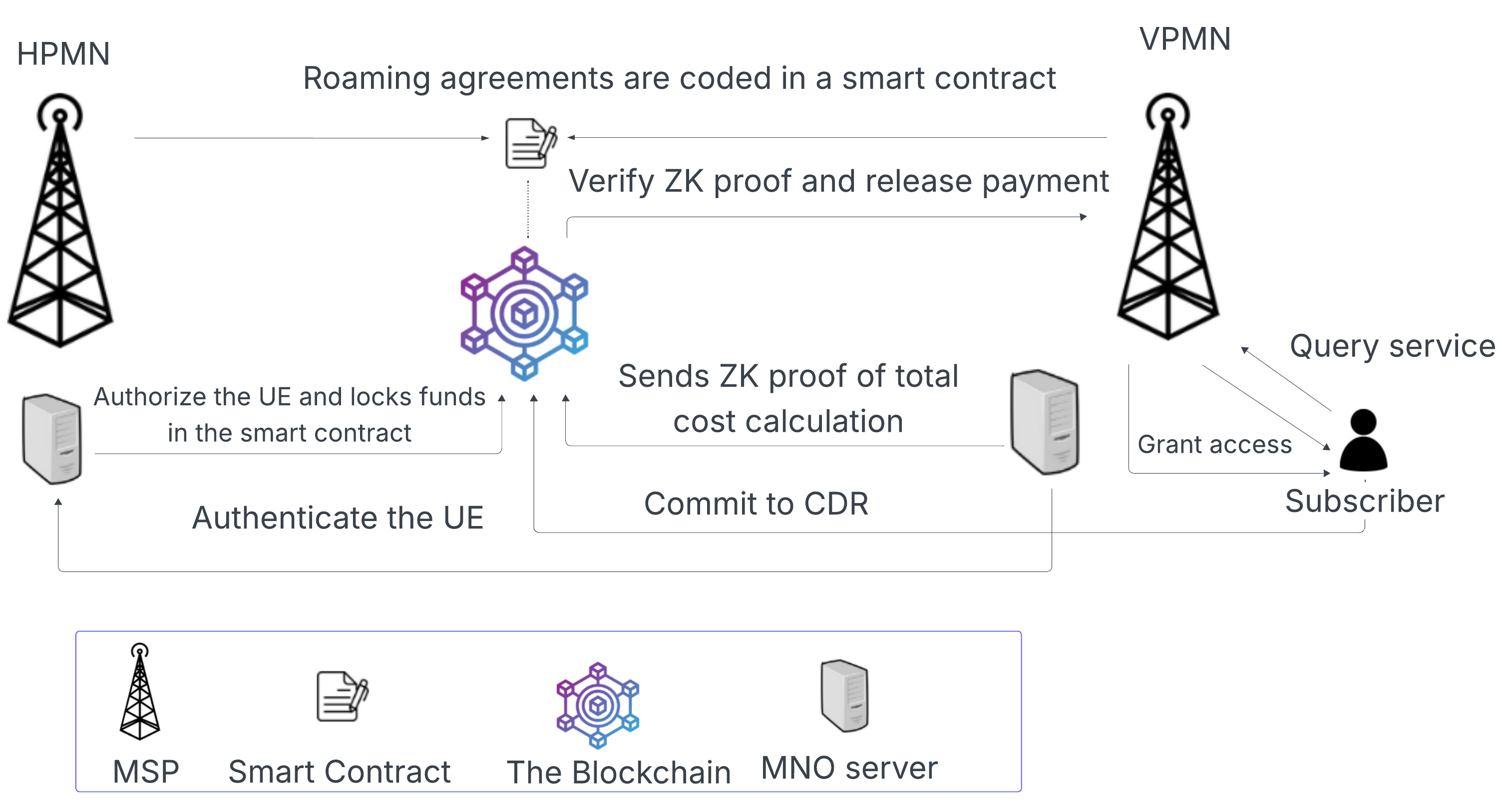}
\caption{System Overview}
\label{fig1}
\end{figure}

\subsection{System Model}

We assume that B5GRoam is deployed on either a public blockchain using a \textit{proof-of-stake} protocol (e.g., Ouroboros \cite{kiayias2017ouroboros}) or a consortium blockchain leveraging \textit{practical Byzantine fault tolerance (PBFT)}\cite{castro1999practical} or one of its variants (e.g., Dynamic PBFT \cite{hao2018dynamic}). As illustrated in Fig.\ref{fig1}, the system involves the following actors:
\vspace{-0.8cm}
\begin{itemize}
    \item \textbf{User Equipment (UE)}. Also known as the subscriber, the User Equipment (UE) refers to the mobile device used by the end user. In B5GRoam, it is assumed that the UE is equipped with a Trusted Execution Environment (TEE). This assumption is well-founded, as most smartphones use ARM-based processors, which integrate TrustZone\footnote{\href{https://developer.arm.com/documentation/100690/0200}{https://developer.arm.com/documentation/100690/0200}}, a widely deployed hardware-based security feature used to isolate and protect sensitive code and data. Furthermore, several manufacturers implement their own TEE solutions on top of this technology, such as TEEGRIS\footnote{\href{https://developer.samsung.com/teegris/overview.html}{https://developer.samsung.com/teegris/overview.html}} by Samsung.
    \item \textbf{Home Mobile Network Operator (HMNO).} It is the subscriber’s original mobile network provider. Its primary role is to authenticate the UE when the user initiates roaming and to authorize the roaming session. The HMNO acts as the payer in the roaming settlement.
    \item \textbf{Visited Mobile Network Operator (VMNO).} It provides the actual roaming service to the subscriber while they are outside the HMNO's network. It is responsible for granting access to the UE and network usage (SMS, data, voice). The VMNO initiates the settlement process and receives payment upon successful settlement.

\end{itemize}

\subsection{Threat Model} In the following, we present our threat model that groups the common attacks a decentralized roaming settlement faces. Assuming the underlying protocol is secure, our goal is to secure the roaming management implemented on the network.
\begin{itemize}
  
    \item \textbf{Sybil attacks}: A common threat in P2P systems/networks where an attacker generates numerous fake identities to gain disproportionate control. Eventually, these \textit{Sybils} are used to disrupt consensus, manipulate reputation systems, or launch further attacks, undermining the system's integrity. \cite{sybill}
    \item \textbf{Poisoning attacks}: Occurs when the VMNO pad or fabricate CDRs usage data volumes, stretching call times, inventing SMS events to make it look as if the UE consumed more resources than reality. The HMNO is then overcharged for phantom usage, and in a decentralized setting there’s no trusted arbiter to catch the manipulation.
    \item \textbf{On-chain privacy leakage attacks}: In decentralized 5G roaming systems, smart-contract billing needs CDR-like data on the public ledger. That on-chain metadata, timestamps, durations, location, and identifiers can be passively scraped by anyone, letting competing MNOs or other observers learn users’ roaming activity and reveal sensitive information.
\end{itemize}

\subsection{Design Overview} \label{subsec:overview}
The B5GRoam framework proposes a Zero Trust architecture for secure, decentralized, and privacy-preserving 5G roaming settlement between MNOs. It integrates zero-knowledge proofs (zkSNARKs) with on-chain verification to ensure the confidentiality and integrity of usage data throughout the settlement process. Unlike traditional roaming systems that rely on centralized intermediaries, such as Data Clearing Houses or IP exchange (IPX) providers, B5GRoam eliminates the need for such entities by automating settlement via smart contracts deployed on a public blockchain with an L2 scaling.

A key improvement over prior works lies in B5GRoam’s treatment of trust assumptions. While all existing blockchain-based solutions assume the VMNO honestly generates CDRs, B5GRoam removes this assumption entirely. Its architecture is designed to remain secure even in the presence of malicious actors by enabling cryptographic verification of CDR correctness without revealing sensitive user data. This trustless design significantly enhances the robustness and applicability of blockchain in real-world 5G roaming. The framework formalizes a roaming agreement between the HMNO and the VMNO as a smart contract on the blockchain. This contract includes predefined billing rates for SMS, data (per MB), and voice calls (per minute). Figure \ref{fig1} illustrates the high-level architecture of B5GRoam, highlighting the different interactions. The process captures the end-to-end roaming settlement workflow, from initial agreement setup to final on-chain payment. \\
\textit{1) Roaming Agreement Deployment}: The HMNO and VMNO negotiate and deploy a roaming agreement as a smart contract on the blockchain. The contract defines unit rates for SMS, data (MB), and voice calls (minutes). \\
\textit{2) Service Query and Subscription}: When a subscriber connects to the visited network, the VPMN receives a service query and prepares to process the roaming request. \\
\textit{3) UE Authentication and Authorization}: The HPMN authenticates the subscriber’s device (UE) and authorizes it for roaming. Simultaneously, the HMNO locks the estimated roaming funds in the smart contract at the beginning of the session. The smart contract thus acts as an escrow, ensuring automatic payment to the VMNO upon session completion without requiring inter-operator trust.\\
\textit{4) Access Grant}: Upon successful authentication, the VPMN grants network access to the subscriber. The subscriber begins roaming, generating usage data for SMS, data, and voice services. \\
\textit{5) CDR Commitment}: After the roaming session concludes, the subscriber’s device generates a cryptographic commitment to the locally recorded usage data. Within the TEE, the device securely collects metrics including the number of SMS messages sent, the volume of mobile data consumed (in megabytes), and the duration of voice calls (in minutes). These values are formatted into a structured CDR, which serves as input to a zkSNARK-compatible hash function. To maintain efficiency and ensure compatibility with zero-knowledge constraint systems, the Poseidon hash function is employed. The resulting hash value representing a cryptographic commitment to the CDR is then submitted to the smart contract. Since the commitment is computed entirely within the TEE, its integrity is ensured, preventing the user from manipulating or forging the committed data.
\begin{equation}
h_\text{cdr} = \text{Poseidon}(n_\text{sms},\ n_\text{mb},\ n_\text{min})
\end{equation}
This hash anchors the CDR on-chain without revealing any sensitive information and acts as a binding commitment that will later be verified against the CDR used in the VMNO's ZKP, enabling trustless validation without revealing the actual usage data. \\
\textit{6) Proof Submission}: Following the roaming session, the VMNO collects its own network-side record of the subscriber's activity, typically in the form of CDRs detailing the number of SMS messages sent, the amount of mobile data used (in MB), and the total voice call duration (in minutes). Based on these values and the rates defined in the smart contract, the VMNO calculates the total billing amount to be charged.

To preserve privacy and enable trustless verification, the VMNO constructs a \textit{zkSNARK proof} attesting that:
\begin{enumerate}[label=\roman*.]
    \item \textit{The computed total fee equals the result of applying the contract's billing rates.}
    \item \textit{The usage metrics hash to the same value previously committed on-chain by the UE, ensuring consistency.}
 \end{enumerate}

The proof is generated using a specific arithmetic circuit that enforces the following constraints:
\begin{equation} \label{equ:cdrcircuit}
\text{total} = n_\text{sms}\!\cdot\!r_\text{sms} + n_\text{mb}\!\cdot\!r_\text{mb} + n_\text{min}\!\cdot\!r_\text{voice}
\end{equation}
\begin{equation} \label{equ:cdrhash}
\text{Poseidon}(n_\text{sms}, n_\text{mb}, n_\text{min}) = h_\text{cdr}
\end{equation}

Only the \texttt{total} and \texttt{committed\_hash} are exposed as public inputs; the actual usage metrics remain private. This enables the smart contract to verify correctness without revealing sensitive user data.

Finally, VMNO submits $(proof, \texttt{total})$ to the Agreement smart contract. \\
\textit{7) On-Chain Verification and Settlement}: The VMNO submits the proof to the smart contract, which verifies the ZKP. Upon successful verification, the contract releases the appropriate payment from the contract to the VPMN, completing the settlement process.

\section{zkSNARK Proof Construction}
In this section, we detail the zkSNARK proof construction in B5GRoam (steps 5 and 6 in Sec.\ref{subsec:overview}). We first introduce its main cryptographic primitives and their properties, then discuss the construction protocol. 
\subsection{Cryptographic Building Blocks} The main cryptographic building blocks upon which the B5Groam system is built are the following:

 \quad \textbf{Hash Commitment.} A cryptographic protocol that allows a party, referred to as the committer, to commit to a chosen value without revealing it, while still being able to prove its validity later on.  A \textit{hash commitment} scheme involves the use of cryptographic hash functions to achieve this goal \cite{grassi2021poseidon}. Let $x$ be the secret value to commit to, $r$ a trapdoor (randomness or key), and $\mathsf{COMM}$ the commitment function. Committing to $x$ involves computing:
\begin{equation*}
   h\_cm = \mathsf{COMM}_r(x) = H(x \| r) 
\end{equation*}
where $H$ is a cryptographic hash function (e.g., Poseidon). Hash commitments are designed to fulfill two properties:
 \begin{itemize}  
   \item \textit{Hiding:} Given $h\_cm$, it should be computationally infeasible to determine the original value $x$.
   \item \textit{Binding:} It should be computationally infeasible to find two distinct values $x_1$ and $x_2$ s.t. $\mathsf{COMM}(x_1) = \mathsf{COMM}(x_2)$.
 \end{itemize}  

\quad \textbf{zkSNARKs.} \textit{Zero-Knowledge Succinct Non-Interactive Arguments of Knowledge} are a type of non-interactive ZKP (NIZK) that enables a prover to convince a verifier of a statement’s validity without revealing any information beyond its correctness. zkSNARKs offer key properties such as zero-knowledge, soundness, completeness, succinctness, and non-interactivity \cite{groth, plonk}.

\quad A zkSNARK (e.g., Groth16) comprises four main algorithms \cite{groth}: 
\begin{itemize} \item $\mathsf{Setup}(1^\lambda)$: Generates public parameters $pp$ from a security parameter $\lambda$. \item $\mathsf{KeyGen}(pp, C)$: Produces a proving key $pk$ and a verification key $vk$ for a computation $C$. \item $\mathsf{Prov}(pk, t, w)$: Generates a proof $\pi$ that the prover knows a private witness $w$ satisfying the computation $C$ on public input $t$. \item $\mathsf{Verif}(vk, t, \pi)$: Verifies the proof $\pi$ against input $t$ and returns true if valid.
\end{itemize}

\subsection{Proof Construction}
We now explain how zkSNARKs are integrated into B5GRoam, outlining the system setup, participating roles, and proof generation process.

\paragraph*{Setup Phase} zkSNARK systems typically require an initial trusted setup phase, commonly known as the \textit{Powers-of-Tau} ceremony. This phase generates structured public parameters, collectively referred to as the \textit{Common Reference String} (CRS), or $pp$, which are used to derive the proving key ($pk$) and verification key ($vk$). These keys are essential for proof generation and verification, respectively, ensuring that both parties operate over the same cryptographic structure tied to the target computation. The setup typically involves multiple independent participants (e.g., MNOs) contributing entropy, which collectively ensures that no single entity can compromise the security of the system. In B5GRoam, this setup is performed once, independently of any specific roaming session.

\paragraph*{Roles in the Protocol} B5GRoam defines two primary roles in the zkSNARK-based billing verification process:

\begin{itemize} \item \textbf{Prover (VMNO):} The Visited MNO constructs a zkSNARK proof demonstrating the correctness of the subscriber’s roaming charges. The proof is computed over private data (detailed usage records) and public data (the commitment hash and the final billing amount).

\item \textbf{Verifier (Smart Contract and External Parties):} Verification is mainly handled on-chain by a smart contract that automatically checks the validity of submitted proofs using the verification key $vk$. Because zkSNARKs are publicly verifiable, any external entity (e.g., HMNO) with access to $vk$ and the public inputs can also verify the proof, without accessing the subscriber’s private data. \end{itemize}

\paragraph*{Proof Generation} After setup and role assignment, the VMNO executes the proof construction process detailed in Algo.\ref{alg:zkproofconstruction}. This process involves evaluating an arithmetic circuit that encodes the billing logic: it checks that the usage data matches the committed hash and that the computed total matches the declared fee. The proof $\pi$ produced by this circuit attests to the correctness of these checks while preserving the privacy of the private data.

\begin{algorithm}[th]
\footnotesize
\DontPrintSemicolon
\SetAlgoLined
\caption{zkSNARK Proof Construction}
\label{alg:zkproofconstruction}
\KwIn{\texttt{n\_sms}, \texttt{n\_mb}, \texttt{n\_min} — usage metrics \\
\hspace{1.5em} \texttt{r\_sms}, \texttt{r\_mb}, \texttt{r\_voice} — billing rates from smart contract \\
\hspace{1.5em} \texttt{h\_cdr} — CDR hash committed by UE \\
\hspace{1.5em} \texttt{total} — Total amount to charge}
\KwOut{\texttt{proof} — zkSNARK proof}

\BlankLine
\texttt{total\_check} $\leftarrow$ (\texttt{n\_sms} $\times$ \texttt{r\_sms}) + (\texttt{n\_mb} $\times$ \texttt{r\_mb}) + (\texttt{n\_min} $\times$ \texttt{r\_voice}) \;
\texttt{h\_check} $\leftarrow$ Poseidon(\texttt{n\_sms}, \texttt{n\_mb}, \texttt{n\_min}) \;
\textbf{assert} \texttt{h\_check == h\_cdr} \;
\textbf{assert} \texttt{total\_check == total} \;
\texttt{circuit} $\leftarrow$ Load ZK circuit with billing + hash constraints \;
\texttt{proof} $\leftarrow$ zkSNARK.Prove(\texttt{circuit}, \\
\hspace{5em} \texttt{private\_inputs} = \{\texttt{n\_sms}, \texttt{n\_mb}, \texttt{n\_min}\}, \\
\hspace{5em} \texttt{public\_inputs} = \{\texttt{total}, \texttt{h\_cdr}\}) \;

\Return (\texttt{proof}, \{\texttt{total}, \texttt{h\_cdr}\}) \;
\end{algorithm}

Specifically, the circuit's constraints ensure that:

\begin{enumerate}
    \item The VMNO's calculated total amount matches the billing computation defined by the smart contract rates.
    \item The VMNO's usage data matches the data usage committed by the subscriber’s UE through the hash commitment.
\end{enumerate}

\noindent By enforcing these conditions, the system ensures both accuracy in billing and privacy of user data, as the zkSNARK proof reveals nothing beyond the correctness of the computation.

\section{Security Analysis} \label{sec:gurusecurity}

\quad We analyze the security of B5GRoam against the threat model introduced in Section~\ref{sec:sysArch}. The framework combines trusted hardware, cryptographic commitments, and ZKPs to mitigate three key attacks:

\textbf{1) Sybil Attacks.}
B5GRoam is deployed on either a public blockchain (e.g., Ouroboros PoS) or a consortium chain (e.g., PBFT or variants like Dynamic PBFT), both of which provide inherent Sybil resistance. In consortium settings, B5GRoam assumes a fixed set of $n = 3f + 1$ validator nodes, tolerating up to $f$ Byzantine replicas. The consensus protocol ensures safety and liveness under standard assumptions (e.g., eventual synchrony). Moreover, each actor—UE, HMNO, VMNO—is uniquely identified by cryptographic keys and interacts via authenticated smart contracts. Since identities are bound to public keys and contractual roles, adversaries cannot gain advantage by creating fake Sybil identities, as these cannot bypass cryptographic verification or influence billing logic.

\textbf{2) Poisoning Attacks.}
To prevent VMNOs from inflating CDRs for profit, B5GRoam enforces verifiable settlement using ZKPs. The VMNO must produce a zkSNARK $\pi$ that proves: (i) The billing amount is correctly computed per the contract-defined formula, and (ii) The CDR used in the computation matches a Poseidon hash previously committed by the UE's TEE. The proof is verified on-chain via the smart contract $\mathtt{Verify}$ using a public verification key $\mathsf{vk}$. Due to deterministic execution and consensus guarantees, all honest nodes accept or reject proofs identically. The soundness and completeness properties of Groth16 ensure the proof cannot be faked or forged, assuming the discrete log hardness in pairing-friendly groups. This eliminates trust in the VMNO's reporting.

\textbf{3) On-Chain Privacy Leakage.}
B5GRoam ensures that no raw CDRs or sensitive metadata (e.g., location, duration) are ever published on-chain. Instead, the UE commits to its usage via a TEE-generated Poseidon hash, leveraging the hiding property of the commitment. The VMNO submits only a zkSNARK proof $\pi$, which attests via the zero-knowledge property, to the correctness of the billed amount with respect to that hash. As a result, even in a fully transparent blockchain environment, passive observers including malicious MNOs cannot infer user activity from billing transactions.

\section{Evaluation \& Results} \label{sec:evals}
\color{black}
\subsection{Experimental Setup}

\quad We develop a proof-of-concept of the B5GRoam\footnote{\href{https://github.com/Brivan-26/B5GRoam/tree/main }{https://github.com/Brivan-26/B5GRoam/tree/main }} framework and run the benchmarks on a Dell XPS 15 9530 laptop with an Intel Core i7 12th-generation processor. We use Hyperledger Caliper\footnote{\href{https://github.com/hyperledger/caliper-benchmarks}{https://github.com/hyperledger/caliper-benchmarks}} to run the evaluation tests and measure on-chain performance using latency and throughput metrics. In addition, we evaluate the performance of different zkSNARK proving systems in the context of B5GRoam. Specifically, we benchmarked three popular proving backends: Groth16\cite{groth}, Plonk\cite{plonk}, and UltraHonk (via Aztec's UltraVerifier). For Groth16 and Plonk, we used the SnarkJS\footnote{\href{https://github.com/iden3/snarkjs}{https://github.com/iden3/snarkjs}} library, while UltraVerifier was evaluated using AztecJS\footnote{\href{https://github.com/AztecProtocol/aztec-packages}{https://github.com/AztecProtocol/aztec-packages}}.

\quad The evaluation is divided into two main parts: (1) Off-chain proving overhead, including proof generation time and memory usage; and (2) On-chain verification overhead, including gas consumption, latency, and throughput.

\subsection{Performance Evaluation}
\quad We consider three metrics for B5GRoam performance evaluation as described below:
\begin{itemize}[leftmargin=0.2cm,align=left]
\item \textbf{Throughput:} Number of successful transactions (TXs) per second (tx/s).
\item \textbf{Latency:} Time interval between TXs submission and its validation.
\item \textbf{Gas:} unit that measures the on-chain computational work of functions.

\end{itemize}

\begin{table}[th]
\centering
\caption{Proving and verification overhead results.}
\label{tab:proof-results}
\begin{tabular}{@{}lccc@{}}
\toprule
\textbf{Proof}                  & \textbf{Groth16} & \textbf{Plonk} & \textbf{UltraHonk} \\ 
\midrule
Generation time ($_{S}$)  & 0.4  & 0.75  & 0.17 \\ 
Generation memory consumption ($_{MB}$)  & 275             & 310 & 90  \\ 
Verfication gas consumption ($_{GAS}$) & 230000             & 270000 & 380000 \\ 
\bottomrule
\end{tabular}
\end{table}

\subsubsection{Off-chain proving overhead}
Table~\ref{tab:proof-results} shows that the three proving systems diverge sharply in both speed and memory footprint, UltraHonk finishes a proof in just 0.17 s, outpacing Groth16 (0.40 s) and Plonk (0.75 s), while using only 90 MB of RAM, roughly one-third of its rivals’ requirements (= 275 MB and 310 MB). These results spotlight UltraHonk as the leading choice for latency-sensitive, resource-constrained deployments such as client-side or off-chain proving, However, fast proof generation must be weighed against the efficiency of on-chain verification.
\begin{figure}[t]
\centering
\includegraphics[width=\columnwidth]{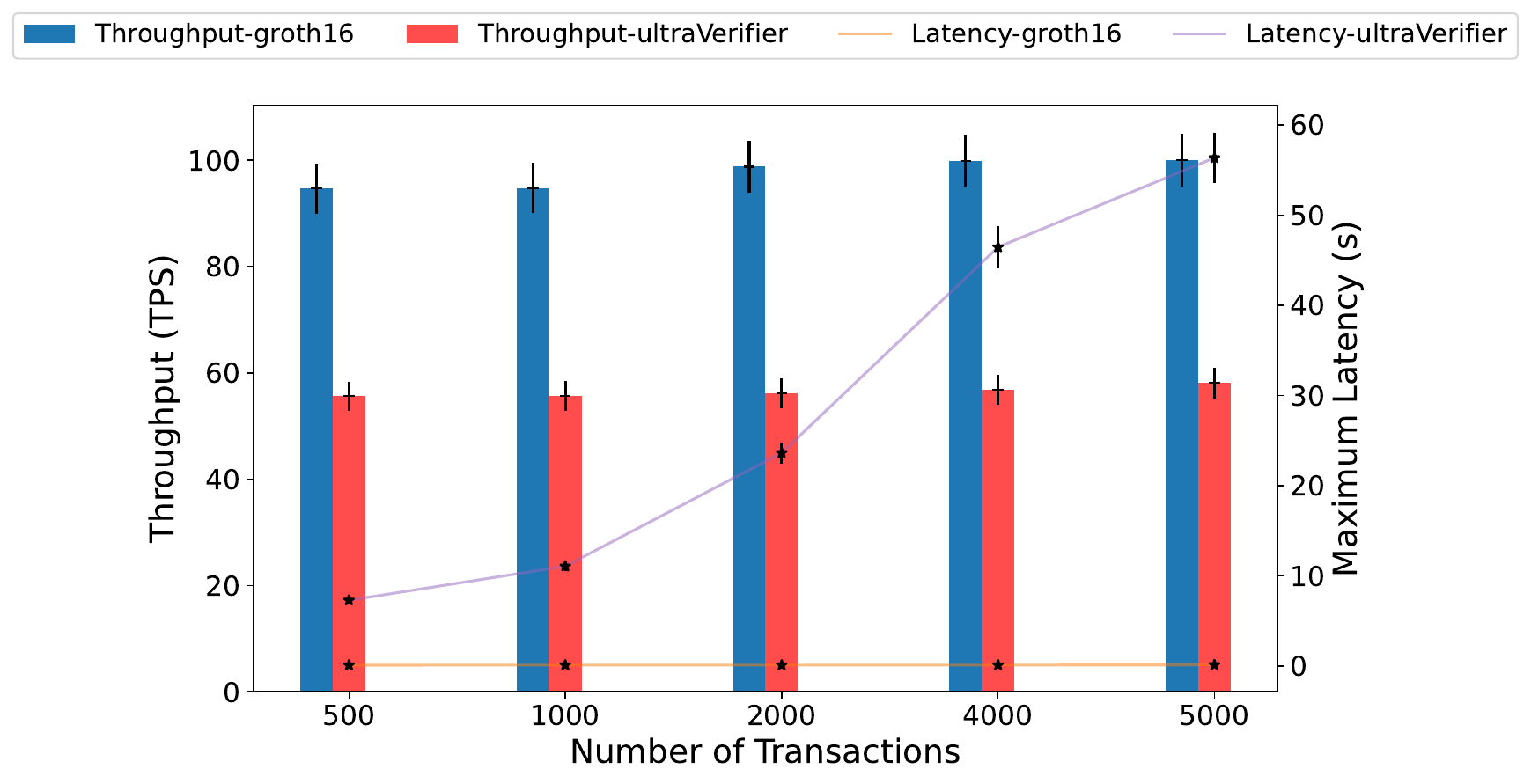}
\caption{L1 throughput and latency comparison.}
\label{fig2}
\end{figure}

\subsubsection{On-Chain verification overhead} 
When verifying proofs on Ethereum, gas efficiency becomes paramount. Among the three systems, Groth16 achieved the lowest gas usage (~230,000 gas), followed by Plonk (~270,000 gas). UltraHonk, despite its speed in proving, incurred the highest verification cost (~380,000 gas) (Tab.\ref{tab:proof-results}). This makes Groth16 the most suitable choice for B5GRoam, where smart contracts and proofs verifications are deployed on-chain. \\
\textbf{Latency:} Stress testing of the verifier under increasing transaction loads \ref{fig2} underscores Groth16's efficiency (i.e., $O(1)$), latency remains close to 0.11 s from 500 up to 5000 proofs, while throughput hovers between 98 – 100 tx/s. UltraVerifier, by contrast, buckles as traffic climbs, latency balloons from 7 s to 56 s, and throughput stalls near 58 tx/s. This shows that Groth16 is well suited to deliver the low-latency, high-throughput pipeline needed for large-scale roaming-settlement workloads.

A critical insight from our benchmark is that the underlying Layer 1 (L1) EVM blockchain may suffice for small to medium-sized roaming workloads. However, it proves too slow and costly for B5GRoam under heavy traffic, resulting in congestion and prohibitively high transaction fees. To overcome these limitations, we opted for Layer 2 zk-Rollups, which are designed to handle a significantly higher volume of transactions, provide near-instant confirmations, and greatly reduce gas consumption, while maintaining security comparable to that of L1.

\subsection{Layer 2 vs Layer 1 Performance}

\begin{table}[h!]
\renewcommand{\arraystretch}{2.5} 
\centering
\caption{Benchmark results comparing Dual Layer and Single Layer transaction costs. Batch size represents the number of L2 TXs within each batch.}
\label{tab:l2_vs_l1_benchmark}
\resizebox{\linewidth}{!}{%
\LARGE
\begin{tabular}{cccccccc}
\hline
\multirow{2}{*}{\textbf{\#Total TXs}}  & \multicolumn{6}{c}{\textbf{Dual Layer 2}} & \multirow{2}{*}{\textbf{Single Layer (L1)}} \\ \cline{2-7}
 & \textbf{ \#Batches} & \textbf{Batch size} & \textbf{Commit} & \textbf{Prove} & \textbf{Execute} & \textbf{Total} &  \\ \hline
60 & 1 & 60 & 205,907 & 83,676  & 99,166 & 388,749 & 15,636,180 \\ \hline
100 & 2 & 50 & 411,802 & 167,328 & 182,794 & 761,924 & 26,060,300 \\ \hline
200 & 3 & 67 & 617,625 & 250,980 & 297,486 & 1,166,091 & 52,120,600 \\ \hline
500 & 7 & 72 & 1,458,233 & 585,588 & 694,162 & 2,737,983 & 130,301,500 \\ \hline
\end{tabular}}

\end{table}

To evaluate the performance benefits of Layer 2 (L2) solutions compared to traditional Layer 1 (L1) execution, we conducted a benchmarking experiment focused on gas consumption of verifying Groth16 proofs on-chain. We used a ZK rollup solution, specifically zkSync\footnote{\href{https://www.zksync.io/}{https://www.zksync.io/}}, and set up a local zkSync environment\footnote{\href{https://github.com/matter-labs/local-setup}{https://github.com/matter-labs/local-setup}} that included both L2 and L1 nodes. Table~\ref{tab:l2_vs_l1_benchmark} presents the results for different transaction volumes.

The results demonstrate a clear and consistent advantage in using L2. For example, processing 60 transactions on L1 consumes over 15 million gas units, while the same workload on L2 costs only around 388,749 units—representing a reduction of over 96\%.

The Layer-2 execution pipeline, comprising $commitBlocks$ (batch formation), $proofBlocks$ (ZKP generation), and $executeBlocks$ (on-chain finalization), adds only modest overhead while delivering a significant cost reduction versus Layer 1. Even when the workload scales to 500 transactions, the total gas consumed on L2 remains far below the L1 baseline. This batching also magnifies throughput: coupling a 60-tx batch size with an underlying Layer 1 capacity of 120 tx/s yields an effective rate of =7,200 tx/s. Collectively, these findings confirm that zk-rollups not only cut fees but also unlock the transaction volume required for real-world deployments.

\section{Conclusion}\label{sec:conclusion}

\quad In this paper, we presented B5GRoam, a blockchain framework that enables secure, scalable, and privacy-preserving roaming settlements for Beyond-5G networks. By replacing trusted intermediaries with zkSNARK-based proofs, it prevents billing fraud and safeguards user data on public chains. Tests with zk-Rollups cut gas costs and push throughput past 7,200 tx/s while maintaining strong privacy. These results lay the foundation for real-world deployment and support the wider adoption of B5GRoam in future mobile networks.

\bibliographystyle{IEEEtran}
\bibliography{ref} 

\end{document}